\documentclass[letterpaper,10pt]{article}
\usepackage{cleveref,citesort,times}
\usepackage{epstopdf}
\usepackage{algorithm}
\usepackage{algpseudocode}
\usepackage{graphics}
\usepackage{epsfig}
\usepackage{subcaption}
\usepackage{booktabs}
\usepackage{tabularx}
\usepackage{amsmath}
\usepackage{authblk}
\usepackage{ucs}
\usepackage{textcomp}
\usepackage{colortbl}
\usepackage{collcell}
\usepackage{hhline}
\usepackage{pgf}
\usepackage{gensymb,siunitx}
\usepackage{etoolbox}
\usepackage{multirow}
\newcommand{\atta}{battery-drain attack}
\newcommand{\attb}{DoB attack}

\graphicspath{{./}}

\begin{document}
%\title{Immobilizing a Car While Its Ignition Is Turned Off} 
\title{\LARGE Who Killed My Parked Car?} 
\date{}

\author[$\dagger$]{Kyong-Tak Cho}
\author[*]{Yuseung Kim}
\author[$\dagger$]{Kang G. Shin}
\affil[$\dagger$]{University of Michigan, Ann Arbor}
\affil[*]{IEEE Member}
\renewcommand\Affilfont{\itshape\small}

\maketitle
\subsection*{Abstract}
%While various ways of attacking and thus controlling the vehicle have been 
%demonstrated, all these attacks were shown to be feasible and effective only
%while the vehicle is running, i.e., ignition is on. That is, the conventional 
%wisdom of vehicle cyber attacks and their defenses is that attacks are 
%feasible 
%and their according defenses are required only when the ignition is on.
We find that the conventional belief of vehicle cyber attacks and their 
defenses---attacks are feasible and thus defenses are required only when the vehicle's ignition
 is turned on---does {\em not} hold. 
We verify this fact by discovering and applying two new practical and important attacks:
{\em battery-drain} and {\em Denial-of-Body-control} (DoB). The former can 
drain the vehicle battery while the latter can prevent the owner from starting 
or even opening/entering his car, when either or both attacks are mounted with the ignition off.
We first analyze how operation (e.g., normal, sleep, listen) modes of ECUs are defined 
in various in-vehicle network standards and how they are implemented in the real world. 
From this analysis, we discover that an adversary can exploit the wake-up function of 
in-vehicle networks---which was originally designed for enhanced user experience/convenience 
(e.g., remote diagnosis, remote temperature control)---as an {\em attack vector}. 
Ironically, a core battery-saving feature in in-vehicle networks makes it 
easier for an attacker to wake up ECUs and, therefore, mount and succeed in 
battery-drain and/or DoB attacks.
Via extensive experimental evaluations on various real vehicles, we show 
that by mounting the {\atta}, the adversary can increase the average battery 
consumption by at least 12.57x, drain the car battery within a few hours or days, 
and therefore immobilize/cripple the vehicle. 
We also demonstrate the proposed {\attb} on a real vehicle, showing 
that the attacker can cut off communications between the vehicle and the driver's key fob
by indefinitely shutting down an ECU, thus making the driver unable to start 
and/or even enter the car.

\section{Introduction}
Software-driven Electronic Control Units (ECUs) and wireless (Wi-Fi,
Bluetooth, Cellular and V2X) connectivities of modern vehicles are proven 
to be double-edged swords.
On the one hand, they have enabled new vehicle applications and services such as
remote diagnosis/prognosis and crash avoidance, enhancing safety, mobility, 
and efficiency. On the other hand, they have introduced more remote 
surfaces/endpoints and thus vulnerabilities which an adversary can exploit and, 
in the worst case, control the vehicle remotely~\cite{attsuf1,checkoway,busoff,attsuf3}.

Researchers have demonstrated how vulnerabilities in remote (e.g., PassThru, 
Bluetooth, Cellular) endpoints can be exploited to compromise an ECU and access 
the in-vehicle network~\cite{checkoway,miller3}.
By exploiting the remotely compromised ECUs, researchers have shown to be 
able to control vehicle maneuvers or even shut down a vehicle via packet 
injection in the in-vehicle network~\cite{teslahack,busoff,koscher,miller}.
%In 2015, researchers were able to compromise and remotely kill a Jeep Cherokee 
%running on a highway~\cite{miller3}, which triggered a recall of 1.4 million 
%vehicles. In 2016 and 2017, researchers were able to hack a Tesla model S and 
%model X, respectively, exploiting software vulnerabilities~\cite{teslahack}.
%%The authors of \cite{miller2} analyzed
%%internal network architectures of 20 different vehicles, and have shown 
%%the practicability and feasibility of remote attacks.
%Researchers also demonstrated that an adversary can also shut down a specific 
%ECU or even the whole in-vehicle network merely via packet 
%injection~\cite{busoff}.
The vulnerabilities that were exploited in (remotely) compromising and thus 
controlling an ECU are, in fact, considered to be inevitable due to the 
inherent nature of automotive 
manufacturing: in-vehicle components and software are developed and written by 
different organizations, and thus vulnerabilities emerge naturally at interface 
boundaries~\cite{checkoway}.
Such a reality of vehicle cyber attacks has made vehicle security one of the 
most critical issues to be addressed by industry, academia, and governments.

While various ways of attacking and thus controlling the vehicle via security 
vulnerabilities have been proposed and demonstrated, all these attacks are 
shown to be feasible and effective, {\em only when the vehicle is running}. 
That is, the conventional belief of vehicle cyber attacks 
and their defenses is that attacks are feasible and their defenses 
are necessary only while the ignition is {\em on}. Thus, there is a lack of 
understanding of what an adversary can achieve or even whether s/he can 
mount malicious attacks while the vehicle's ignition is {\em off}. 

In this paper, we show such a general belief does {\em not} hold since 
an adversary can attack and control a parked vehicle (i.e., with ignition 
off) and immobilize it indefinitely. Specifically, we propose two new attacks---% 
{\em battery-drain} and {\em Denial-of-Body-control} (DoB)---which make 
the vehicle inoperative.

By mounting the battery-drain attack, an adversary first {\em gains} access to the in-vehicle network
then {\em controls} various functions of the car, and finally {\em drains/discharges} the 
battery to a level where the car cannot be started. 
In this paper, we refer to ``immobilize'' as an action that prevents the driver 
from starting or driving the car.  
As the ignition is off, one might think that no matter what message(s) the attacker injects,  
none of the in-vehicle ECUs would receive, and respond to, the injected messages. 
Surprisingly, however, our analyses of various in-vehicle network 
standards and their protocol implementations reveal the feasibility of controlling ECU 
functions via message injections even when the ignition is  {\em off}. 
Ironically, the main reason for this feasibility is the ``wake-up functions''---which are intended to 
enhance the driver's experience/convenience---let the adversary wake up ECUs (of a parked 
vehicle) and then control them. That is, the wake-up functions that were 
originally designed for a good cause become an {\em attack vector}. 
Wake-up functions are standardized, implemented, and thus provided in various 
in-vehicle networks so that car manufacturers can provide remote standby 
functions, such as remote diagnostics, door/temperature control, and anti-theft. 
Without the wake-up functions, the ECUs providing such functionalities 
would have to run continuously, hence consuming too much of battery.

Therefore, exploiting such a standardized and thus available wake-up 
function, the attacker (i) wakes up ECUs by injecting a 
wake-up message, (ii) controls the awakened ECUs by sending certain messages 
(e.g., those that turn on lights, unlock/lock the door, change power mode, open trunk, etc.), 
and therefore, (iii)  achieves his goal of draining the vehicle's battery.
In order to control such functions, the attacker must know which messages 
(more specifically, with which message IDs) to inject, usually requiring some 
(painstaking) reverse-engineering with fuzzing~\cite{checkoway,busoff}. 
However, for the purpose of  {\atta}, we propose a driver-context-based 
scheme, which significantly lowers the technical barrier for the adversary to
reverse-engineer the required control messages, i.e., figuring out which 
message IDs to use, thus helping the attacker succeed in {\atta}.

Through the proposed {\em Denial-of-Body-control} (DoB) attack, in addition to 
simply waking up ECUs (as was also done in the {\atta}), an adversary can force 
all awakened ECUs to enter the ``bus-off'' state, i.e., shut-down. The 
attacker does this is to exploit the fact that, depending on 
their software configuration, some ECUs do {\em not} recover from a bus-off; a 
policy specified in the ISO 11898-1 standard~\cite{iso}.
We find through evaluations on real vehicles that a DoB attack can, in fact,
lead to a case where important ECUs, such as a Remote Control Module (RCM)---%
which is an integral part of remote key and security functionalities---do not 
recover from a bus-off, i.e., remain shut down. As a consequence, the 
communication between the key-fob and the RCM (i.e., vehicle) is cut off, thus 
making the driver unable to enter or start his vehicle.

%For simplicity and more importantly for low-power design (thus having 
%a prolonged battery lifetime), such a wake-up message/signal is defined in the 
%standards to be very simple (e.g., one 0-bit).
%From the attacker's perspective, this implies that the wake-up message can 
%easily be fabricated. Once the attacker wakes up ECUs, it then controls 
%certain 
%functionalities of the vehicle (e.g., lock/unlock the door) via message 
%injection and finally 

It is important to mention that the proposed battery-drain and DoB attacks are 
interesting, critical, and very different from the attacks known to date for the 
following reasons.

\begin{itemize}
	\item One common irony of the two proposed attacks is that the 
	wake-up function of in-vehicle networks, which was originally standardized, 
	designed, and built for enhanced user experience/convenience, is exploited as 
	an {\em attack vector}.  Capitalizing on the wake-up function, the adversary 
	becomes capable of mounting the attacks even while the ignition is off.
	
	\item The feasibility of, or ease in mounting the attacks stems from the 
	fact that the wake-up signal/message was defined to be very {\em simple}.
	A simple (agreed-on) wake-up message (e.g., one 0-bit) facilitates the design
	of low-power ECUs/transceivers, thus extending the  battery operation time. 
	We refer to ``battery operation time'' as how long the battery can last to provide enough 
	power for the driver to start the car. 
	From a security viewpoint, however, such a battery-saving feature makes it 
	easier for the attacker to wake up ECUs and then drain the vehicle's battery.
	
	\item The simplicity of the wake-up signal makes message encryption or use of Message 
	Authentication Codes (MAC) (some state-of-the-art defenses) unable to prevent an attacker 
	from waking up ECUs.
	
	\item The number of ECUs that can/must be awakened, tends to increase 
	as more enhanced standby features are added to newer vehicle models. 
	This will allow the attacker to immobilize the newer models far more 
	quickly and easily than the older ones.
\end{itemize}

Through extensive experimental evaluations of 11 real vehicles%
--- i.e., 2008--2017 model-year (compact and mid-size) sedans, 
coupe, crossover, PHEV (Plug-in Hybrid Electric Vehicle), SUVs, truck, and an 
electric vehicle---we show that all (except one 2008 model-year) of our test 
vehicles are equipped with the wake-up functions, rendering both battery-drain and DoB attacks feasible. 
Moreover, we show that by mounting a {\atta}, the adversary can speed up the 
average battery consumption by at least 12.57x, drain the car battery within a 
few hours, and therefore immobilize the vehicle. 
We also demonstrate the proposed {\attb} on a real vehicle and show 
that the attacker can shut down an ECU indefinitely and thus prevent the 
driver from entering or starting the car.

In summary, we make the following main contributions:
\begin{enumerate}
\item Showing the feasibility of waking up ECUs via message injections by 
	analyzing in-vehicle network protocols, standards, and implementations, and
	demonstrating it on 11 different real vehicles;
	%\item Proposal of a new driver-context-based reverse engineering method 
	%which 
	%lowers the technical barrier for an attacker mounting the {\atta}; and
\item Discovery of two new attacks---battery-drain and DoB attacks---through which an 
adversary can immobilize vehicles while the ignition is off; and

\item Demonstration of the two newly proposed attacks on a real vehicle.
\end{enumerate}

%The rest of the paper is organized as follows. Section~\ref{sec:primecan} 
%provides the
%required background on CAN and related work. Section~\ref{sec:volfing}
%details the threat models under consideration while Section~\ref{sec:volfing} 
%details \name.
%Section~\ref{sec:evaluation} evaluates \name\ on a CAN bus prototype
%and on real vehicles, and
%Section~\ref{sec:disc} discusses the various aspects of \name.
%Finally, we conclude in Section~\ref{sec:conclusion}. \]

\section{Background}\label{sec:primecan}
\subsection{Terminal Control}
The car battery powers in-vehicle ECUs not only when the 
ignition is on but also when it is off.
When and how much battery/power an ECU drains depends on the {\em 
terminal} it is connected to. In the DIN 72552 standard~\cite{dinstandard}, 
terminals are defined as follows. Connected ECUs attached to
\begin{itemize}
	\item Terminal 15: switched on with ignition on and (totally) off when 
	ignition is off
	\item Terminal 30: permanently powered on but usually runs in sleep 
	mode, while the vehicle is parked and locked (i.e., ignition is off).
\end{itemize}
This definition of differentiation in terminal control is to provide different 
functionalities in vehicles when the ignition is on/off.
As an example, consider the Passive Keyless Entry 
and Start (PKES) system, which allows users to open and start their cars while 
keeping their car keys in their pockets~\cite{Francillon11relayattacks} and is 
equipped in most contemporary vehicles. In order to provide the keyless entry 
feature, the ECU running PKES will have to be connected to terminal 30 and be 
permanently powered on, continuously sensing whether the owner's/driver's key 
fob is within a certain range.
Meanwhile, if the power modes of such permanently powered on ECUs are not 
properly managed, they will quickly drain the car battery.
Thus, to minimize their power consumption, as described in the DIN 72552 
standard, they operate in {\em sleep} mode in which only their transceiver (not 
their microcontroller) is powered on.
This is for the ECU's transceiver to still be capable of detecting and 
therefore responding to any wake-up signals. 
For this reason, transceivers have a separate power supply~\cite{wakeuptutorial}.
This way, while the ignition is off, ECUs asleep switch to, and 
operate in {\em normal} mode, i.e., wake-up, only when required, thus preventing
fast drain of the battery and typically keeping cars continuously parked/idle for 25$\sim$40 days
without losing battery power.
The battery doesn't get charged by the alternator until the vehicle engine runs back again with 
the ignition on.

\begin{table}[!t]
	\footnotesize
	\centering
	\begin{tabular}{cccc}
		\toprule
		\textbf{Bus} & \textbf{Data Rate} & \textbf{Industry Standard} & 
		\textbf{Wakeup} \\
		\midrule
		\multirow{3}{*}{CAN} & 500 kBit/s &
		\begin{tabular}[c]{@{}c@{}}ISO 11898-1\\ ISO 11898-2/5\end{tabular} & 
		Global \\\cline{2-4}
		& 125 kBit/s & \begin{tabular}[c]{@{}c@{}}ISO 11898-1\\ ISO 
			11898-3\end{tabular} & Global\\\cline{2-4}
		& 33.3 kBit/s & GM LAN & Global\\\hline
		LIN & 19.2 kBit/s & LIN Consortium & Global\\\hline 
		\begin{tabular}[c]{@{}c@{}}SAE\\ J2602\end{tabular} & 10.4 kBit/s &
		SAE J2602 & Global\\\hline
		FlexRay & \begin{tabular}[c]{@{}l@{}}10 MBit/s \\ 5 MBit/s \\ 2.5 
			MBit/s\end{tabular} & FlexRay Consortium & Global \\\hline
		\begin{tabular}[c]{@{}c@{}}MOST\\(Multimedia)\end{tabular} & 
		\begin{tabular}[c]{@{}c@{}}100 MBit/s \\ 1 GBit/s 
		\end{tabular} 
		& MOST Cooperation & -  \\
		\bottomrule
	\end{tabular}
	\caption{{\em Overview of vehicle bus systems with wake-up 
			capability~\cite{wakeuptutorial2}. }
		%https://www.infineon.com/dgdl/Infineon-Energy_Saving_in_Automotive_EE_Architectures-White-Paper-WP-v01_00-EN.pdf?fileId=5546d462525dbac40152d4b618cb0063
		In all in-vehicle networks, but MOST, ECUs are woken up
		simultaneously, i.e., a global wake-up, when a wake-up signal is detected 
		on the bus.}
	\label{tab:wakeup}
\end{table}

\subsection{Wake-up}
Even when the ignition is off, the need of ECUs asleep to be awakened
is increasing for enhanced user/driver experience/convenience. 
For example, vehicle OEMs are providing useful functions, such as PKES,  
overnight remote diagnostics, remote temperature control, remote door control, 
and anti-theft while the vehicle is parked and turned off.
We refer to such functionalities that are executable/executed while the ignition is off 
as {\em standby} functions.
To meet such a need, ECUs asleep are designed and configured to be 
awakened via two different mechanisms: local wake-up and bus wake-up.

A local wake-up is triggered when a switch attached to the ECU (e.g., a receiver 
for the remote key) is turned on. This drives a logic state change on its WAKE 
pin and thus re-activates the whole ECU.
Another mechanism in which an ECU wakes up is whenever it sees a specific 
in-vehicle network signal, i.e., wake-up message/signal, on the bus. 
Upon detection of a wake-up signal by the ECU's transceiver, which remains ON even 
while the ignition is off, it turns on the power supply of the ECU, thus waking 
up the whole ECU, i.e., enters normal operational mode. 
We will later elaborate on what the {\em wake-up signals} are when we 
discuss the proposed attack methods.
If no additional wake-up signal is received within a certain (preset) period, 
the ECU returns to sleep mode.

Table~\ref{tab:wakeup} summarizes in-vehicle network standards/protocols that define such 
a wake-up functionality/capability of providing a pathway for driver-friendly applications.
One can see that it is standardized, implemented, and used not only in CAN---the {\em de facto} 
standard of in-vehicle networks---but also in all other in-vehicle networks, except for MOST. 
However, for the purpose of more in-depth discussion, we focus on CAN when discussing 
the two proposed attacks.

\begin{figure}[!t]
	\psfull \centering
	\includegraphics[width=0.7\linewidth]{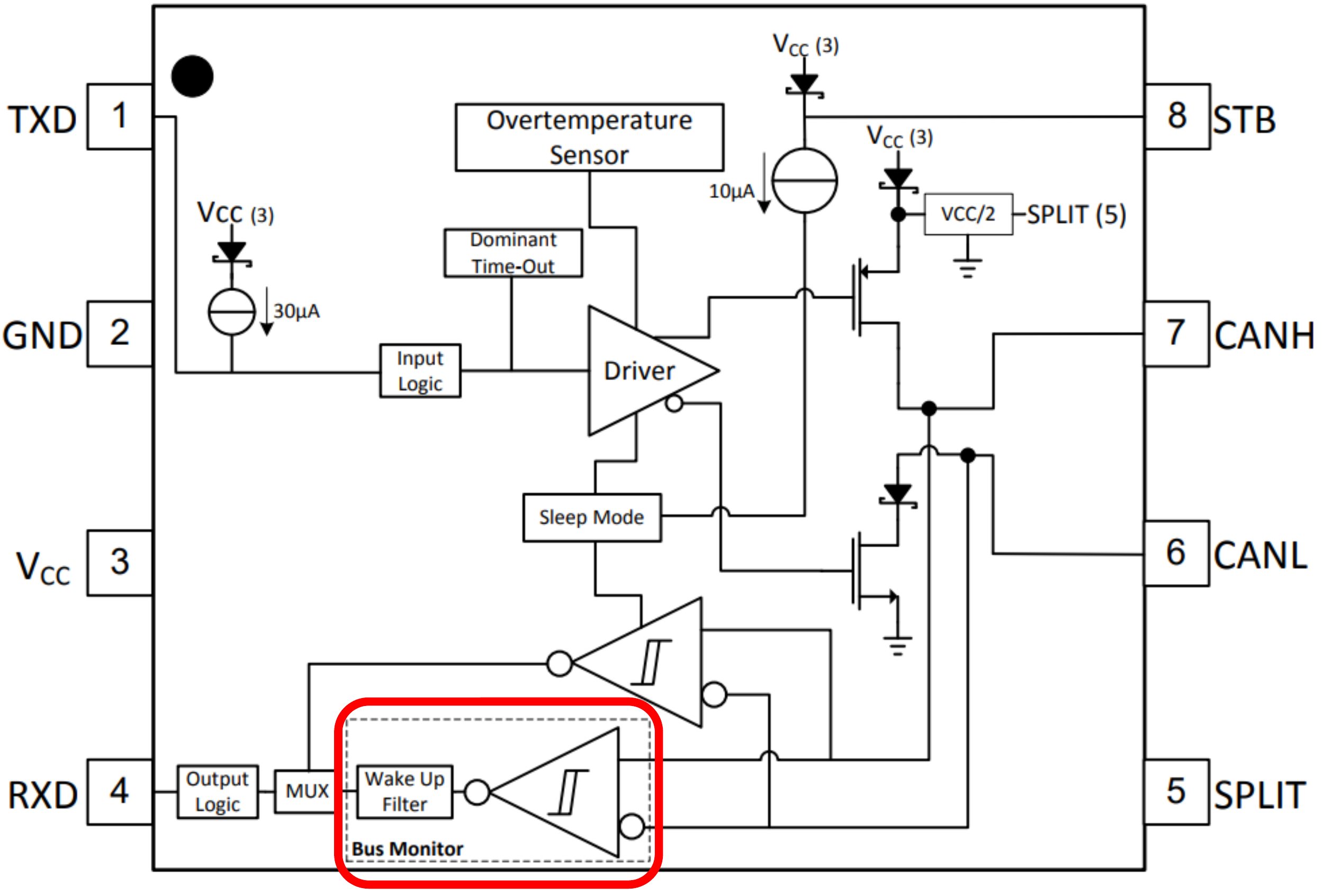}
	\caption{{\em TI SN65HVD1040 transceiver~\cite{TIwakeup}.} 
	The bus monitor module continuously  checks whether or not there are any wake-up signals on the bus. 
	If so, it wakes up the whole ECU.}
	\label{fig:TIwakeup}
\end{figure}

CAN ECUs/transceivers with such a wake-up feature (i.e., wake-up detection module) have been 
in the market for more than 15 years (as of 2017)~\cite{wakeup_infineon}. 
Fig.~\ref{fig:TIwakeup} shows a block diagram of a Texas Instrument (TI) SN65HVD1040 transceiver 
implemented with a bus monitor module for the wake-up function. We found that (almost) all CAN 
transceivers from other manufacturers also have such a wake-up detection module 
with different names like wake-up detector, wake-up filter, etc.

%%%%%%%%%%%%%%%%%%%%%%%%%%%%%%5%% Do we need this?
%\subsection{Body Control Module (BCM)}
%The Body Control Module (BCM) is a processor-based ECU (or body computer) that 
%supervises and controls functions related to the car's body such as lights, 
%windows, security, door locks, and various comfort controls.
%%%%%%%%%%%%%%%%%%%%%%%%%%%%%%%%%%%%%%%%%%%%%%%%%%%%%%%%%%%%%%%%%%%%%%%%%%%%%%

\section{Adversary Model}\label{sec:attackmodel}
We consider an adversary whose objective is to {\em immobilize} the victim's vehicle while 
the ignition is off, and stay as {\em stealthy} as possible  during, and even after the attack.
By immobilizing the vehicle, i.e., compromising its availability, s/he prevents 
the driver from starting or driving the car, % via ignition control. 
making the vehicle unavailable when the driver/owner wants or needs to use it.
In order to be as stealthy as possible during, and even after the attack, the 
adversary also aims to immobilize the car {\em before} the 
driver attempts to enter/start it. Therefore, although an adversary may in 
fact immobilize a vehicle by simply flooding the in-vehicle network when the 
driver attempts to start it---which, in turn, prevents any other ECUs (e.g., 
Body Control Module) from receiving/processing commands---or by 
jamming the key-fob wireless channel, we do not consider such an adversary. 
Immobilizing the vehicle in such ways is likely to expose the attacker,
%when attacking the vehicle, 
since the driver/owner would be in the vicinity of the car when the attack is mounted. 
In contrast, by immobilizing the vehicle before the 
driver attempts to enter/start it, the adversary won't leave any
trace at all, except the immobilized vehicle, i.e., very stealthy.

%Since different vehicles have different types of software and hardware, they 
%would have different attack surfaces and thus degrees of 
%vulnerabilities. Hence, depending on the victim's vehicle, the approach of the 
%adversary in immobilizing it might have to be different.
%More importantly, the best effort for the adversary in achieving his goal  
%would heavily depend on his capabilities and knowledge. 
%For example, based on whether the adversary can physically/remotely compromise 
%an in-vehicle ECU (on the CAN bus) through some attack surface and mean, his 
%methodology of immobilizing the vehicle would be different. 
%Therefore, we consider the following two different
%types of adversaries --- {\em CAN} attacker and {\em Key-fob} attacker --- who 
%have different capabilities and attack means.

%\subsection{CAN attacker}
As in previously discussed attacks~\cite{checkoway,koscher,busoff,miller3},
we consider the adversary capable of remotely (but {\em not physically}) compromising 
an in-vehicle ECU via numerous attack surfaces and means, and can thus gain its 
control; physically compromising an ECU requires physical access and thus is 
not stealthy.
That is, the adversary can compromise 1) a (third-party) OBD-II 
(On-Board Diagnostics) dongle/device in advance, and gain remote control of 
the vehicle once the driver plugs it in to his/her car (as demonstrated in 
\cite{woot_tele})\footnote{Once it is plugged in, we can consider the OBD-II 
dongle/device as an in-vehicle ECU.} or 2) an in-vehicle ECU (e.g., telematic 
unit which has external connectivities), remotely, so as to access the 
in-vehicle network~\cite{checkoway,miller3}.
Since such an adversary would have access to the vehicle's CAN 
bus, we call such an adversary a {\em CAN attacker}.
Once an ECU is compromised, we consider the CAN adversary to be
capable of performing at least the following malicious actions. The
adversary can inject any message with forged ID, DLC (Data Length Code), and 
data---which we refer to as {\em attack messages}---on the bus as they are 
managed at user level. 
Also, since CAN is a broadcast bus, the adversary can sniff messages on CAN. 
These are the basic capabilities of a CAN adversary who has the control of a 
compromised ECU. The practicability of such an adversary model has already
been proved and demonstrated in~\cite{checkoway,koscher,miller}.

%
%\subsection{Key-fob Attacker}
%We also consider a different type of adversary who --- in contrast to the CAN 
%adversary --- does {\em not} have access to the victim's CAN bus; either due 
%possibly to lack of his technical expertise or resource. On the other hand, we 
%consider the adversary to have the capability of either 1) compromising the 
%victim's key fob, 2) designing a malicious key fob himself, and/or 3) 
%developing a special (key-fob wireless channel) jamming device.
%Since such an adversary exploits a key fob and/or its communication channel  
%as 
%his attack vectors for (victim's) vehicle immobilization, we refer to such an 
%adversary as a {\em Key-fob} attacker.
%
%\kt{need more detail for this.}

\section{Immobilizing a Vehicle}\label{sec:immobilize}
We now introduce two new attacks through which a CAN adversary can
immobilize the victim's vehicle while the ignition is off, i.e., compromise the
vehicle's availability while it is parked and turned off.

\begin{figure}[!t]
	\psfull \centering
	\includegraphics[width=0.7\linewidth]{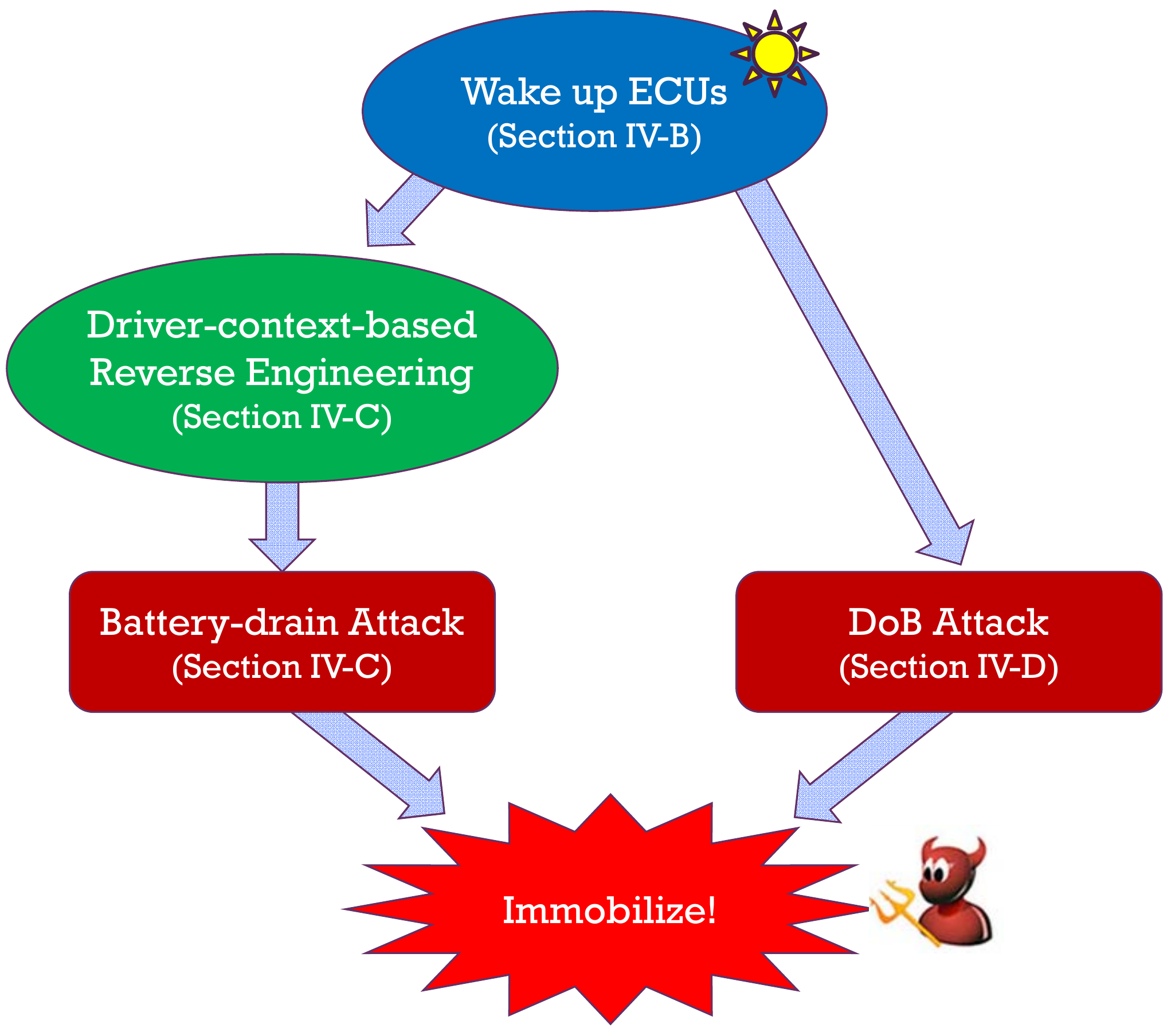}
	\caption{{\em An overview of the proposed attacks.}
		After waking up ECUs in a parked vehicle, the CAN adversary
		immobilizes the vehicle via a {\atta} and/or a {\attb}.}
	\label{fig:overview}
\end{figure}

\subsection{High-level Overview of the Proposed Attacks}
Through a compromised in-vehicle ECU (e.g., telematic unit, OBD-II dongle),
the CAN adversary
has access to the CAN bus irrespective of whether the ignition is on or off.
However, especially when the ignition is off and thus all ECUs are
asleep or turned off (until the ignition is turned on again), an attack message
injected by the adversary may not be delivered to those ECUs.
That is, no matter what messages the adversary may inject, these ECUs
may not respond.

Fig.~\ref{fig:overview} shows an overview of how the CAN adversary immobilizes
a vehicle in such a case. In order to control/attack ECUs on the bus, s/he first
wakes them up and then immobilizes/cripples the vehicle via a
battery-drain or DoB attack. In the case of battery-drain attack
(Section~\ref{sec:batterydrain}), the adversary will
attempt to exploit the awakened ECUs for controlling certain functionalities of the
vehicle (e.g., illuminating exterior/interior lights) and thus draining its battery.
In order to figure out which message ID to use for such a control,
the attacker goes through a message reverse-engineering
process based on {\em driver context}, which will be detailed in
Section~\ref{sec:reverse}. That way of reverse-engineering messages
allows the attacker to succeed in {\atta} much easier than via
conventional reverse-engineering such as fuzzing.
In case of {\attb} (Section~\ref{sec:dobattack}), the attacker need not go
through the
reverse-engineering process,  because the {\attb} does not control ECUs but
shuts them down by exploiting their error handling and recovery mechanisms.
By mounting a battery-drain and/or a DoB attack, the attacker immobilizes the
vehicle.

\subsection{Waking Up ECUs}
When the ignition is off, some ECUs are configured to run in sleep mode and
continuously monitor if there is any incoming local or bus wake-up signal.
Taking this into consideration, using his/her compromised ECU, the CAN
adversary attempts to wake up all other ECUs by delivering a {\em bus wake-up} signal
to them. Then, what type of {\em bus wake-up signal} should the adversary use in
order to wake up the ECUs?

{\bf Standardized wake-up signal.}
The remote wake-up behavior of a CAN ECU was first introduced and specified
in the ISO 11898-5:2007 standard which defines the bus wake-up behavior as:

\vspace{0.2cm}
{\em ``One or multiple consecutive dominant (0-bit) bus levels for at
least $t_{Filter}$, each of them separated by a recessive (1-bit) bus 		
level, trigger a bus wake-up.''}
\vspace{0.2cm}

\noindent The ISO standard specifies $t_{Filter}$ to be within [500ns, 5$\micro$s].
We can thus deduce the following three facts:
\begin{itemize}
	\item [\bf S1.] Dominant bus level (i.e., bus with 0-bits) for longer than
	5$\micro$s causes a wake-up;
	\item [\bf S2.] Dominant bus level for shorter than 500ns is ignored; and
	\item [\bf S3.] Dominant bus level for a period between 500ns and
	5$\micro$s
			{\em may} cause a wake-up.
\end{itemize}

\begin{figure}[!t]
	\psfull \centering
	\includegraphics[width=\linewidth]{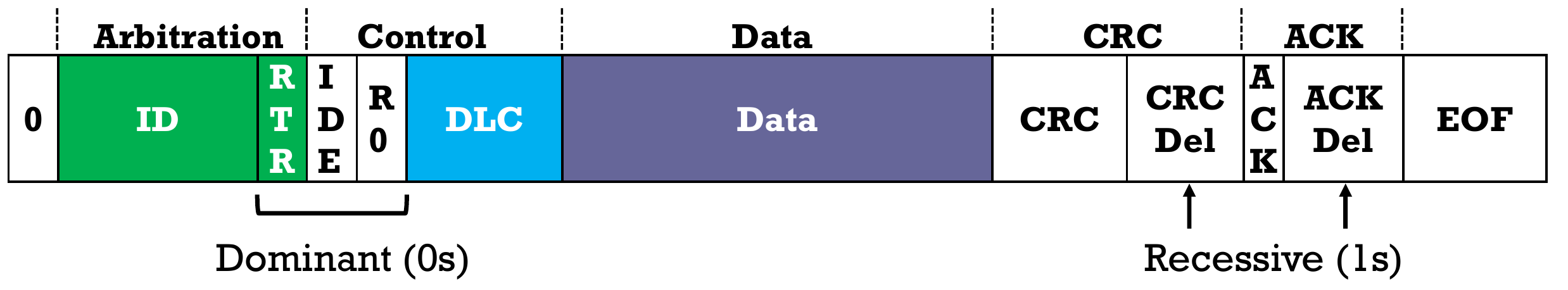}
	\caption{{\em Format of a CAN data frame.}
		For a normal CAN data message, RTR, IDE, and r0 bits are defined to be 0s
		and CRC and ACK delimiters are defined to be 1s.}
	\label{fig:canframe4}
\end{figure}

\noindent The actual value of $t_{Filter}$ depends on the transceiver being used.
In a CAN bus with bit rates up to 200 kBit/s, i.e., bit width longer than 5$\micro$s,
the dominant bus level condition S1 is met with any 0-bit within a CAN frame/message.
%Such a CAN bus is usually referred to as CAN-2, i.e., Medium se and normally
%interconnects ECUs responsible for vehicle body control.
For bit rates up to 500 kBit/s, %i.e., {\em CAN-1},
the dominant condition S1 is also met for any normal CAN data message since its
1) RTR, 2) IDE, and 3) r0 bit (ISO11898-1:2003) (later revised to an FDF bit in
ISO11898-1:2015) are all defined to be dominant (0s) as shown in Fig.~\ref{fig:canframe4}.
That is, in a 500 kBit/s bus, since those three bits---each with
width 2$\micro$s---are sent consecutively, the resulting duration of dominant bus level
becomes at least 6 $\micro$s, thus (automatically) guaranteeing/satisfying S1 to be met.

Note that the ISO standard specifies such dominant bus levels to be
separated by a recessive bus level. This is easy to achieve since CAN always
1) stuffs a recessive 1-bit after 5 consecutive 0 bits, i.e.,
bit-stuffing~\cite{canspec,busoff},
2) has certain fields fixed with a 1-bit (e.g., CRC delimiter, ACK delimiter
as shown in Fig.~\ref{fig:canframe4}), and 3) the user can determine
what value(s) to fill in such fields as ID, DLC, and DATA.
The same also applies to the extended CAN format which has a 29-bit ID.
However, in this paper, we only consider the basic/standard CAN data format
since the extended
format is seldom  used (due to its bandwidth waste) in contemporary vehicles;
most vehicles use the basic/standard format with 11-bit IDs.

The reason for OEMs' agreement on a standardized wake-up signal was to
guarantee a 100ms link acquisition time~\cite{wakeuptutorial}. More
importantly, the
the wake-up signal was defined to be {\em simple} to allow for a
low-power design (e.g., RC-circuit) of wake-up detection, i.e., an
energy-efficient sleep mode~\cite{wakeuptutorial}. OEMs want to reduce the
average standby/asleep current consumption to be less than 300$\micro$A per
ECU, and many of them reduce it even further down to less than
100$\micro$A~\cite{nxpbattery}.
That is, the simple design/definition of a wake-up signal was to prolong the
battery operation time, i.e., energy-efficient.\footnote{Energy-efficiency of vehicles is no longer
an option, but is a prerequisite when defining and developing new ECUs. From 2020 onwards,
there will be a very challenging threshold for CO$_2$ emissions with 95g CO$_2$/km for
passenger cars sold in Europe~\cite{wakeuptutorial2}.}

Similarly to CAN, other in-vehicle networks such as FlexRay and LIN define the
wake-up signals to be simple for energy-efficiency. FlexRay specifies the
signal to be long high/low~\cite{wakeuptutorial} and LIN
specifies it to be a dominant bus level within the interval [250$\micro$s,
5ms]~\cite{lin}. The wake-up signal in those networks also wakes up all
ECUs asleep on the bus; see Table~\ref{tab:wakeup}. An adversary can, therefore, easily wake up ECUs
while the ignition is off, in not only CAN bus but also other in-vehicle networks. However, in this paper,
we focus on CAN %, the {\em de facto} standard in-vehicle network,
for a more in-depth treatment of the attacks.

{\bf CAN adversary waking up the bus.}
As the wake-up signal is simple and standardized, all the CAN adversary
needs to do is inject a fabricated {\em wake-up message/signal} into the bus,
which s/he has access to. As mentioned earlier, due to its simple
definition, a wake-up message with {\em any} content (i.e., ID, DLC, and DATA
which are controllable by a remote attacker) will wake up ECUs.
Note, however, that only those ECUs which are asleep (i.e., not completely
off) will be awakened.
This is ironic/paradoxic since wake-up signals are made simple for energy-efficiency,
i.e., to minimize battery consumption, but such a simple design helps the attacker
wake up ECUs, thus making the vehicle {\em less} energy-efficient.

{\bf Power source of the adversary's ECU.}
One remaining requirement for a CAN adversary to achieve this is that
his (compromised) ECU has to remain powered on.
The attacker achieves this fairly easily thanks to two interesting facts.

First, the ECUs which a CAN adversary would (or can) compromise and thus
use are most likely to be continuously powered on, or (at least) have a
separate power source/supply during their operation.
A typical example ECU that an attacker would target (to compromise) is the telematic unit
due to its wide variety of external/wireless connectivities. The practicability of the telematic
unit being compromised has already been demonstrated in \cite{checkoway,koscher,miller}.
Interestingly, the telematic unit---which is regarded as one of the most
vulnerable ECUs~\cite{busoff,checkoway,woot_tele,cids}---is usually completely (or at least periodically)
powered on so as to respond to external events (e.g., requests for remote diagnosis,
remote door/temperature control, anti-theft) even after the ignition key has been taken out~\cite{teleon}.
Moreover, a telematic unit is usually equipped with an alternative power
supply so that it can operate even when the car battery or electrical system
is faulty~\cite{telematicpower}.
Similarly, an OBD-II device/dongle, which is also a good target for an adversary
to compromise (as demonstrated in~\cite{checkoway,woot_tele}), can also be completely
powered on (by the attacker) since it is also normally equipped with an external
power source (e.g., battery).
In summary, ECUs which a CAN adversary will likely compromise
via their exposed attack vectors are the ones which are completely/always
powered on either by the car battery or their own power supplies.

\begin{figure*}[!t]
	\psfull \centering
	\includegraphics[width=\linewidth]{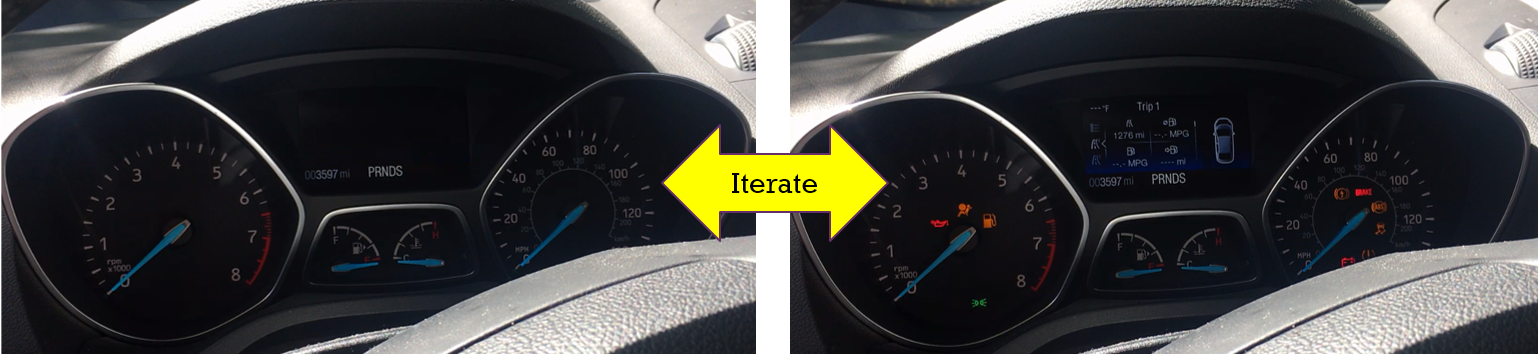}
	\caption{{\em Consequence of controlling the power mode of a parked
	vehicle.}
		After waking up ECUs (asleep in a parked vehicle), when we periodically
		injected messages with an ID that corresponds to changing the power
		mode of the test vehicle, various indicator lights continuously
		illuminated/flickered even though the ignition was off.
		% After a short period, those lit-up indicators disappeared, unless it
		%was illuminated/controlled periodically.
	}
	\label{fig:powermode}
\end{figure*}

Second, although the operational mode of a (compromised) ECU is preset to
run in sleep mode when the ignition is off, it does not restrict the adversary
to change such a setting. Two most common CAN controllers---%
Microchip MCP2515 and NXP SJA1000---both allow modification of their operation
mode (e.g., normal, sleep, listen-only) through software
commands~\cite{mcp2515,sja1000}. For ECUs with the Microchip MCP2515 CAN
controller, the Serial Peripheral Interface (SPI) remains active even when the
MCP2515 is in sleep mode, thus allowing access to all registers. Thus,
through the SPI, it is also possible for the user/adversary to read/write the
CAN controller registers, including the operational mode
register~\cite{mcp2515}. Such user-level features for configuring the CAN
controller allow attackers to easily switch from sleep to normal mode via
software commands.

As a result, a CAN adversary can inject wake-up messages to the CAN bus
while the ignition is off. The transceivers of ECUs asleep observe a wake-up signal
on the bus, switch on the ECUs' power supply (usually via an interrupt), and boot up the
microcontroller. Hence, the ECUs return to normal operational mode.
Since CAN is a broadcast bus, even a single injected message from the CAN
adversary causes all ECUs asleep to run in that mode.

\subsection{Battery-Drain Attack}\label{sec:batterydrain}
We now give a detailed account of how an adversary can drain the battery of a vehicle
via message injections, i.e., mount a {\atta}. Here, we only give the details
of how the {\atta} can be mounted. Section~\ref{sec:evaluation} will detail
their consequences via in-depth evaluations on real vehicles.

\subsubsection{Attack 1 --- Waking up ECUs}
While the ignition is off, the CAN adversary can first attempt to wake up ECUs
via message injections. Once the ECUs asleep wake up, they switch to, and run in
normal operational mode. Note, however, that an awakened ECU goes back to
sleep after a certain period of time (configured by the OEM).
Hence, by waking up ECUs as much and as frequently as possible, the adversary can
{\em continuously} force those ECUs to run in normal mode, although they should remain asleep.
If ECUs are configured to stay in normal mode (after waking up) for a
duration of $T_{wakeup}$, the frequency of wake-up messages from the attacker has
to be at least $\frac{1}{T_{wakeup}}$. Since the adversary can read/sniff
messages on the CAN bus, s/he can easily infer $T_{wakeup}$ from the monitored traffic.

With continuous injections of wake-up messages, ECUs (that would usually be asleep) will
be forced to stay up in normal mode and thus draw much more current (i.e., power) from the battery.
In contrast to the ECUs asleep, which normally consume less than 100$\micro$A of
the battery current, normal-mode ECUs consume several mA.
Therefore, by simply waking up ECUs---the simplest {\atta}---the CAN
adversary can significantly increase the battery current consumption and can
thus reduce the battery operation time.
We will later in Section~\ref{sec:evaluation}
detail the amount of current drained by simply waking up ECUs, and how that
affects the vehicle battery operation time.

\subsubsection{Attack 2 --- Controlling ECUs}
An interesting consequence of waking up ECUs is not only the increased battery current consumption
but also the pathway it provides for the attacker to {\em control} ECUs.
After waking up, since ECUs previously asleep now run in normal operational mode---the same as
when the ignition is {\em on}---the CAN adversary becomes capable of controlling them.
We refer to ``controlling an ECU'' as executing the ECU's function(s) via message injections.
For example, an attacker may inject an attack message with ID=0x11, which is
usually sent by some other ECU. If message 0x11 is processed and used
by the brake ECU in engaging/disengaging the brake, an injection of 0x11 will
control that ECU's brake function and thus the corresponding vehicle maneuver.

However, the criticality levels of some (malicious) controls would be different
from when the vehicle's ignition is on and moving, compared to the case when the
vehicle is parked with its ignition off, i.e., different from existing attacks.
For example, ensuring that the brakes do not unwillingly engage/disengage in a
moving vehicle is safety-critical. It might not be critical when the ignition is off and the vehicle is parked.
From the {\em battery/energy consumption} perspective, the controls
that would be considered {\em malicious} are different.

{\bf Controls that increase battery consumption.}
Since the activation of interior/exterior lights is one of the highest
battery-consuming functionalities, the adversary can attempt to control them to
increase the battery current drain. %, and (probably) deplete it.
Instead of attempting to {\em directly} control the interior/exterior lights
(via light-control messages), we exploit the following vehicle functions which
{\em indirectly} illuminate various lights inside and outside the vehicle.
\begin{itemize}
	\item[\bf C1.] Changing the vehicle's power mode;
	\item[\bf C2.] Repeatedly unlocking and locking doors; and
	\item[\bf C3.] Opening the trunk.
\end{itemize}
The reason why we exploit such indirect functions is that their control
messages are far easier to reverse-engineer (i.e., figure out the
meaning/purpose of messages) than the direct (light-)control messages if done
based on {\em driver context}.\footnote{To control vehicle functions, we must
know which message ID(s) to use/inject. However, since that
information is OEM-proprietary, we must reverse-engineer them.}
This fact counters the common belief that reverse-engineering
CAN messages is a non-trivial painstaking process~\cite{busoff}.
We will later in Section~\ref{sec:reverse} detail how the message reverse-engineering
process can be eased, especially for the purpose of {\atta}.

{\bf C1. Changing the vehicle's power mode.}
Depending on the ignition switch position, vehicles run in different power
modes such as off, accessory mode, run, and start. An adversary exploits this fact
and can first reverse-engineer the control message (via driver context) which
determines/reflects the vehicle's power mode, wakes up ECUs on the bus, and
attempts to change the power mode using the reverse-engineered message/ID.

Fig.~\ref{fig:powermode} shows why changing the power mode of a vehicle can be
considered as an indirect way of illuminating lights and thus increasing the
battery consumption, i.e., mount a {\atta}.
It shows what happened when we tried to alter the power
mode of one of our test vehicles via message injections. Here, every
procedure and consequence was executed and happened while the ignition was
off. We  will later in Section~\ref{sec:evaluation} provide the evaluation settings
and general methodologies in accessing the CAN bus and injecting messages.
Note that it was fairly easy to reverse-engineer (or figure out) the power-mode
control message, especially when using the context---a main reason why we
attempted to mount the {\atta} via power-mode control instead of (direct)
light control. When we controlled the power mode of a parked vehicle, as shown
in Fig.~\ref{fig:powermode}, various indicators on the dashboard were
(temporarily) illuminated.
Similarly, albeit not shown in Fig.~\ref{fig:powermode}, the infotainment
system was also booted up.
When we {\em periodically} injected the reverse-engineered ``power-mode
control'' message to the bus, the indicators on the dashboard continuously
flickered and the infotainment system was intermittently turned on.

Although the vehicle ignition was off, the injection of a (fabricated) power-mode control message
was successful since the vehicles' power-mode master is the
Body Control Module (BCM)---the ECU which has to be at least asleep (but never
completely off) due to its important role in providing various standby functions.
BCM not only plays a central role in a car by maintaining control over
its various functions such as power windows, air conditioning, and central
locking, but also governs security functions including Remote
Keyless Entry (RKE), Vehicle Anti-Theft Security System (VTSS), and Passive
Anti-Theft System (PATS).

In summary, if a CAN adversary not only wakes up ECUs but also
reverse-engineers and injects power-control messages, s/he can continuously
illuminate various lights/indicators and further increase the
overall battery consumption: an enhancement of {\atta} that simply wakes up ECUs.

{\bf C2. Unlocking and locking the door.}
In addition to the previous two types of {\atta} --- waking up
ECUs and controlling the power mode---a CAN adversary can also attempt to
repeatedly unlock and lock the vehicle (while parked). The reason why a CAN
adversary would mount a {\atta} in this way is not only 1) the unlock/lock messages
are easy to reverse-engineer but also 2) it activates various light functions.
When the driver (or the adversary) unlocks the car, {\em welcome lights} of the
vehicle illuminate for an enhanced visibility for the driver. Note that the
numbers and types of the welcome lights that illuminate may vary with
the vehicle manufacturer/year/model, and also depending on whether the lighting
control system is set as ``automatic,'' which is the default setting for most
drivers~\cite{checkoway}.

During daytime, lights such as marker lights, interior lights, exterior footlights
(on side mirrors), and coming-home lights will/might illuminate when the driver
unlocks the vehicle. At night time, the vehicle will turn on its headlights.
By reverse-engineering and injecting ``door lock control'' messages to the bus and then exploiting
these {\em driver-friendly lighting controls}, an adversary can continuously illuminate all the welcome
lights and thus significantly increase the average battery consumption; another enhancement of {\atta}.
Similarly to the attack case C1 where the power mode was controlled, the door
control module is another ECU which has to provide a standby function such as
keyless entry and must thus be not completely off (i.e., must be asleep instead).
This allows the attacker to control the door locks even when the ignition is off.

{\bf C3. Opening the trunk.}
The adversary can also attempt to open the trunk of a car similarly to the
attack cases C1 and C2. Again, the trunk control module/ECU is another ECU
that would have to be asleep, not completely turned off. Therefore, by
reverse-engineering the trunk control message, an adversary can (remotely) open
the trunk. When the trunk is opened, for enhanced visibility for the driver,
(almost all) vehicles are configured to illuminate its interior map, dome, and
trunk lights. Again, thanks to such user-friendly lightings, the adversary can
illuminate more lights inside/outside the vehicle and thus further increase the
battery consumption, i.e., reduce the battery operation time.

In contrast to C1 and C2, the attacker is only required to
inject a single trunk-control message into the bus if the lights remain on
while the trunk is open (as some vehicles do).
Even if the lights automatically go off after some time, the attacker
can re-inject the trunk-control message to re-illuminate them.
Note, however, that opening the trunk could be a bit (visibly) intrusive, which
is a limitation of this approach, although for some vehicles we observed that
it is not. However, if the adversary can deplete the battery before the
driver/passenger notices and thus attempts to close it, such
an intrusive approach will still succeed in immobilizing the vehicle.
When mounted overnight, since the driver/passenger would notice this only when
s/he attempts to start the car in the following morning, the attacker could be given approximately
half a day or even a few days (e.g., weekends) in succeeding it!

\begin{figure}[!t]
	\centering
	\includegraphics[width=0.7\linewidth]{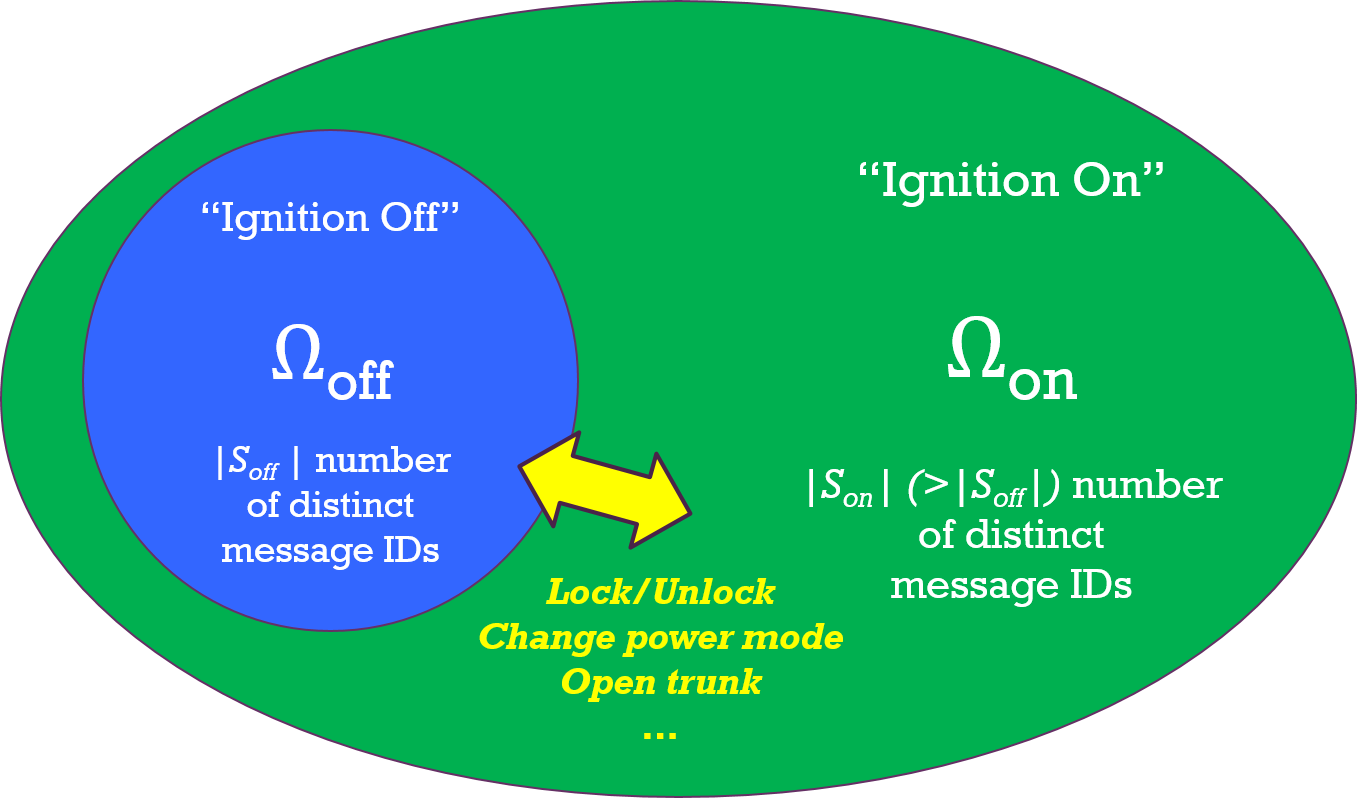}
	\caption{{\em Driver-context-based reverse-engineering.} Using this
		approach, a CAN adversary can easily figure out which message IDs to
		use in mounting the {\atta}.}
%		 controls such functions as door lock/unlock, change in power mode,
%		and opening the trunk because the driver usually (or even always)
%		makes such actions before starting the car.}
	\label{fig:context}
\end{figure}

\subsubsection{Driver-Context-Based Reverse-Engineering}\label{sec:reverse}
In controlling an ECU (as in C1--C3), since the message IDs that have to be
used are different for different vehicle manufacturers and models, adversaries
would have to reverse-engineer messages for each vehicle that it targets.
This fact generally becomes a (high) technical barrier for the adversary,
especially when mounting state-of-the-art attacks on different vehicles.
However, for the purpose of {\atta}, the message
reverse-engineering can be achieved very differently, i.e., not via fuzzing.
Specifically, by reverse-engineering messages based on the {\em
driver-context}, it becomes much easier for the adversary to figure out which
messages to use in mounting and thus succeeding in {\atta} on
different/various vehicles.

The proposed driver-context-based reverse-engineering works as follows.

\begin{enumerate}
	\item When the ignition is off, the CAN adversary continuously wakes up
	ECUs and records/logs the CAN traffic as $\Omega_{off}$ as illustrated in
	Fig~\ref{fig:context}.
	If the adversary knows when the driver usually starts the car (e.g., 9am in
	the morning), s/he can start such a process just before that time.
	We define the sets of distinct IDs sent while the ignition is on and off as
	$S_{on}$ and $S_{off}$, respectively. Then, once ECUs are awakened via a
	wake-up message from the adversary, $|S_{off}|$ distinct message
	IDs would be observed on the bus and that number would be lower than
	$|S_{on}|$, since the number of ECUs running in normal operation mode
	(and thus periodically sending messages) would be less.
	
	\item The CAN adversary continuously logs the CAN traffic as $\Omega_{off}$
	and marks the {\em bit positions} of messages' ($\in S_{off}$) data fields
	that continuously change as ${\Delta}_{off}$.
	One example of ${\Delta}_{off}$ can be the last byte of the data field where OEMs
	usually put their 1-byte checksum of each message (ID)~\cite{busoff,miller3}.
	
	\item When the CAN adversary finds that the ignition is (now) turned
	on, onwards, s/he records/logs the traffic (of an ``ignition on''
	vehicle) as $\Omega_{on}$. The CAN adversary can acknowledge this
	by observing a suddenly-increased number of distinct message IDs---%
	from $|S_{off}|$ to $|S_{on}|$---on the bus.
	
	\item Since $S_{off} \subset S_{on}$, the CAN adversary analyzes how the
	data values/fields, excluding ${\Delta}_{off}$, of those message IDs
	($\in S_{off}$) changed during the period of just before to right after the
	ignition being started.
	%between before and after the ignition started.
\end{enumerate}

An interesting yet important fact about $S_{off}$'s data values (excluding
those in ${\Delta}_{off}$) is that they reflect the driver's {\em actions} before driving the vehicle.
Imagine a person who tries to start and drive his/her car.
S/he would first unlock the vehicle and then change its power mode, i.e., turn on the ignition.
Perhaps, s/he might even open the trunk to put items there before starting the vehicle.
Such a routine of accessing and starting the vehicle, which we define as the {\em driver context},
is  embedded/reflected in $\Omega_{off}$, i.e., the CAN traffic obtained while the ignition was off.
Specifically, when the driver makes some action (e.g., unlock the
door) while the ignition is off, the data value of an ID ($\in S_{off}$) would change.
Note that such a change would not incur in any of the data fields in
$\Delta_{off}$, i.e., the bit positions which their values normally continuously change,
since the driver's action and thus the corresponding change in the data values are ``temporary''.
So, by observing the temporary data changes in $S_{off}$ incurred right before the ignition is turned on,
the CAN adversary can easily figure out which messages relate to those driver-context-related controls.
Interestingly (and luckily for the CAN adversary), as discussed in
Section~\ref{sec:batterydrain}, such driver-context-related controls lead to illuminating various indicators/lights.
As a result, especially for a CAN adversary attempting to mount a
{\atta}, reverse-engineering the required control messages (e.g., door
lock/unlock, changing power modes) becomes far more easier!
The same can be applied when the driver stops and leaves the vehicle, since
s/he would again change the power mode, perhaps open the trunk, and of course,
lock the vehicle.

Given below is an example of how we reverse-engineered one of our test
vehicle's door-control messages. When describing the procedures, we do not use
the actual ID value nor the bit positions, since they are proprietary
to the OEMs. When comparing two sets of $\Omega_{off}$, one before
pressing a remote key fob and another during it, we found that the data
fields of message ID=0x01 had changed from
%\begin{center}
[00 \underline{10} 00 00 FF 00 AB CD]
%\end{center}
%\noindent
temporarily to
%\begin{center}
[00 \underline{30} 00 00 FF 00 BC EF].
%\end{center}
We verified in advance that the last 2 bytes of message 0x01
continuously change, even without any actions taken on the vehicle.
That is, we verified that the last two data bytes of 0x01 belong to
$\Delta_{off}$. As a result, we were able to easily figure out that the
second byte of message 0x01 controls our test vehicle's lock and unlock
functions. We will later show through evaluations that such an approach was indeed
successful and thus let us easily unlock \& lock the car, illuminate the
welcome lights, and therefore, mount the {\atta}.

\begin{figure*}[!t]
	\centering
	\begin{subfigure}[h]{0.35\linewidth}
		\centering
		\includegraphics[height=4.3cm]{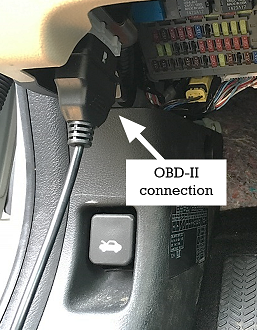}
		\caption{Messages were injected to the parked test
			vehicle through the OBD-II port.}
		\label{fig:obdconn}
	\end{subfigure}
	~
	\begin{subfigure}[h]{0.63\linewidth}	
		\centering
		\includegraphics[height=4.3cm]{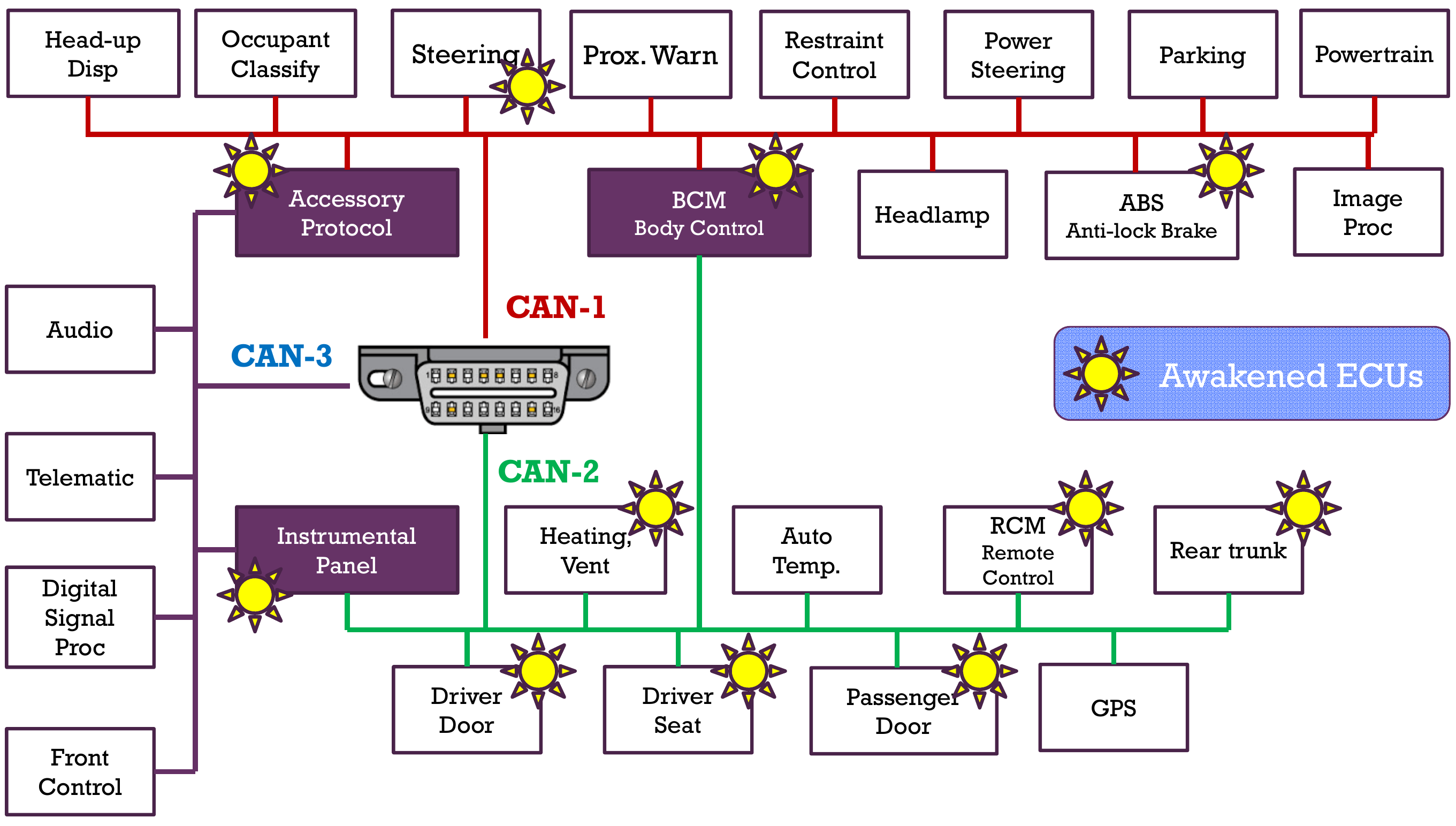}
		\caption{Vehicle architecture of one of our test vehicles. Once a
		wake-up message was injected, several ECUs were awakened.}
		\label{fig:ford_wakeup}
	\end{subfigure}
	\caption{{\em Analysis of how our test vehicle responded to a wake-up
			message.} We established connection to our test vehicle's
			CAN bus through the OBD-II port and examined which and how many
			ECUs woke up when a wake-up message was injected.}
\end{figure*}

\subsection{Denial-of-Body-control (DoB) Attack}\label{sec:dobattack}
In addition to the {\atta}, the CAN adversary can mount a Denial-of-Body-control (DoB) attack
in order to immobilize a vehicle, i.e., compromise its availability.

{\bf Bus-off recovery.}
Error handling is built in the CAN protocol and is important for
its fault-tolerance. It aims to detect errors in CAN frames and enables
ECUs to take appropriate actions, such as discarding a frame,
retransmitting a frame, and raising error flags.
If an ECU experiences or incurs continuous errors while transmitting or receiving a message,
the CAN protocol specifies that its Transmit Error Counter (TEC) or Receive Error
Counters (REC) should be increased, respectively~\cite{canspec}.
If its TEC exceeds a pre-defined threshold of 255 due to continuous errors, the
ECU is forced to enter a state called {\em bus-off} and shut down.

Exploiting such a standardized CAN feature, Cho and Shin~\cite{busoff}
proposed a new attack called the {\em bus-off attack}, which enforces other
healthy/uncompromised ECUs to shut down.
See \cite{busoff} for more details of the bus-off attack.
The proposed {\attb} is mounted in a similar way to the bus-off attack,
except that it further exploits the following fact specified in the ISO 11898-1
standard~\cite{iso} and thus immobilizes the vehicle. The standard
specifies that

\vspace{0.2cm}
{\em ``A node can start the recovery from the bus-off state only upon a user's
request.''}
\vspace{0.2cm}

\noindent where the user's request depends on the ECU's software/policy
configuration. The proposed {\attb} thus exploits such a definition of
bus-off recovery as follows.

While the ignition is off, the CAN adversary wakes up ECUs so as to make them
responsive to his injected messages. Then, the adversary switches
its bit rate (e.g., from 500 kBits/s to 250 kBits/s).
According to the CAN error-handling scheme, this makes {\em all} awakened
ECUs on the bus continuously experience and incur errors, and therefore enter
the bus-off state, i.e., shut-down.
This way, the adversary not only mounts the bus-off attack on a targeted ECU
(as demonstrated in \cite{busoff}) but also on all ECUs on the bus.
Instead of changing the bit rates, changing internal/net resistances or
capacitances can be an alternative method in achieving this.
As a result, per bus-off recovery specification, depending on the ECUs'
software configurations, some ECUs would recover from the bus-off state, whereas
some will {\em not}, i.e., remain shut down.

Depending on the car manufacturer and year/model, ECUs such as BCM or RCM,
which is the security ECU that authenticates each message to and from the remote key
fob, can in fact be configured/defined not to recover from the bus-off state,
mainly for safety, since the bus-off is a serious problem~\cite{busoff}, or for
anti-theft purposes.
Hence, if the CAN adversary were to mount the {\attb} on such a vehicle,
then s/he can indefinitely shut down the BCM or RCM, and thus cut off the
communication between the (driver's) remote key fob and the vehicle.
Contemporary/newer vehicles are mostly equipped with the PKES system, which
allows users to open and start their cars while
having their key fobs in their pockets~\cite{Francillon11relayattacks} and is
installed either in BCM or RCM. For the vehicle to be opened/started, PKES must
verify that the legitimate key fob is in the vehicle's vicinity.
Therefore, shutting down BCM/RCM (and thus PKES) would mean that the vehicle
will {\em not} be able to receive and authenticate any remote key signals (sent
by the key fob), thus preventing  opening or starting the vehicle, i.e., the CAN
adversary immobilizes the vehicle via a {\attb}.

Once the attacker succeeds in mounting {\attb}, there is no need for the
attacker to mount the attack, again; some ECUs that have entered bus-off will never
boot up again, anyway.
This allows the attacker to not only succeed in mounting the attack in a very
short period of time but also leave without any trace, i.e., stealthy, except the immobilized vehicle!
We will later show through evaluations that such a configuration of BCM/RCM not
recovering from bus-off actually exists in real vehicles and thus makes
the driver unable to open the door/trunk and start the vehicle even with
his/her legitimate, perfectly-functioning key fob.

\section{Evaluation}\label{sec:evaluation}
We now evaluate the feasibility and criticality of the two proposed attacks---% 
{\em battery-drain} and {\em DoB} attacks---on various real vehicles.

\begin{figure*}[!t]
	\psfull \centering
	\includegraphics[width=\linewidth]{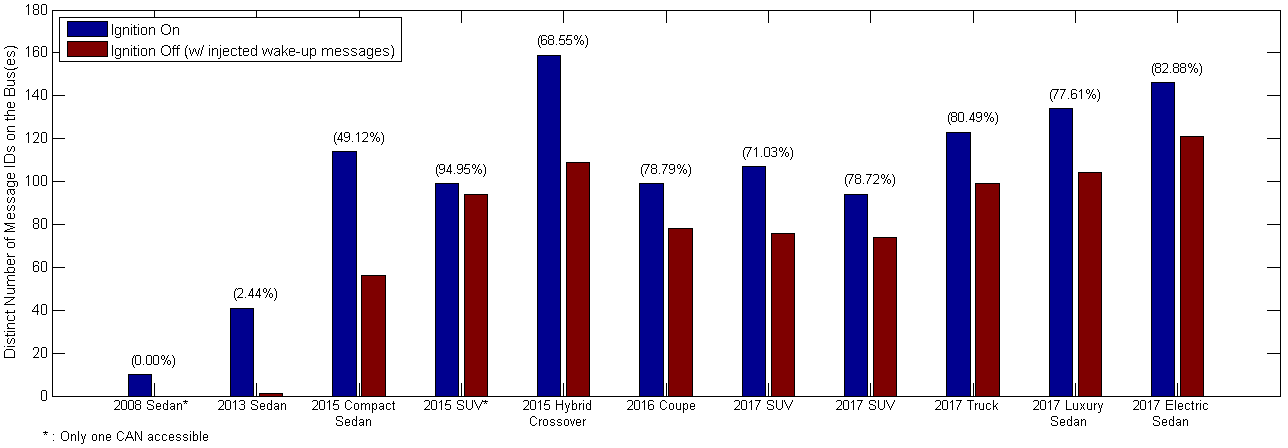}
	\caption{{\em Verifying wake-up messages in 11 different vehicles.}
		We compared how many distinct message IDs were observed when the 
		ignition was off but the ECUs were awakened via a wake-up message and 
		when the ignition was on.}
	\label{fig:wakeup_eval}
\end{figure*}

\subsection{Waking Up ECUs}
To verify whether ECUs in real vehicles can indeed be awakened via 
simple wake-up messages while the ignition is off, we connected a Vector CAN 
device to our test vehicles' OBD-II port as shown in Fig.~\ref{fig:obdconn}.\footnote{The 
OBD-II port can be used to access all CAN buses in a vehicle %This is a standard function since 
and is thus the principal means service technicians use to 
diagnose and update individual ECUs~\cite{checkoway}.}
We then injected wake-up messages/signals to the CAN bus.
The wake-up message we used had its ID, DLC, and DATA fields --- that 
a user/adversary can control at the application layer --- all filled with 1s. 
This was to verify that messages with the minimum number of 0s can also 
properly function as a valid wake-up message.

{\bf In-depth analysis on a test vehicle.}
Fig.~\ref{fig:ford_wakeup} shows the in-vehicle network architecture of one of 
our test vehicles --- a 2017 year model\footnote{The model identity is not 
revealed for the OEM's confidentiality.} 
--- and the ECUs that responded to the wake-up signals/messages. 
Note that this network architecture of the test vehicle is not unique 
for the OEM of our test vehicle but is general/valid for the vehicles built by other OEMs; 
there are only slight variations in the network architecture~\cite{miller2}.
We verified which ECUs were awakened by logging the CAN traffic that contains 
the message IDs observed on the bus, and by mapping those IDs to the 
corresponding transmitter ECUs using the test vehicle OEM's CAN Data Base 
Container (DBC). The CAN DBC describes the properties of the CAN network, 
the ECUs connected to the bus, and the CAN messages and signals.
For the purpose of this research, the CAN DBC file was provided by 
the test vehicle's OEM.\footnote{Due to its proprietary information, such DBC 
files are not shared without permission from the OEMs and their suppliers.}

When the wake-up message/signal was injected, one can see from 
Fig.~\ref{fig:ford_wakeup} that not all ECUs were awakened. 
This would be most probably due to different ECUs being attached 
to different terminals (with different terminal control policies) in 
order to minimize battery/power consumption and, at the same time, provide 
various standby functions. In CAN-1 which connects ECUs performing safety-critical 
functions, only 4 of 13 of them were awakened. On the other 
hand, in the CAN-2 bus where ECUs responsible for vehicle body control were 
connected, almost all ECUs but two were awakened. Considering the fact that 
contemporary/newer vehicles provide various standby ``body control'' functions 
such as keyless entry, hands free (foot) trunk opening, and anti-theft, more 
number of ECUs being awakened in CAN-2, i.e., bus with body-control ECUs, than 
CAN-1 would be the norm.

\begin{table*}[!t]
	\centering
	\scriptsize
	\begin{tabular}{lccc}
		\toprule
		Attacks               & Discharged Current {[}mA{]} & Amplification 
		Factor 
		& Max. Battery Operation Time {[}Days{]} \\
		\midrule
		None                 & 12.2                       & 
		Baseline             
		& 30.7                        \\
		+ Wake-up             & 42.0                       & 
		x3.44                   
		& 8.92                        \\
		+ Change Power Mode   & 74.5                       & 
		x6.11                 
		& 5.02                           \\
		+ Lock \& Unlock Door & 101.1                      & 
		x8.29                  
		& 3.70                        \\
		+ Open Trunk          & 153.3                      & 
		x12.57                  
		& 2.44                        \\
		\bottomrule
	\end{tabular}
	\caption{{\em Maximum battery operation time under different battery-drain 
			attacks.} Based on the measured battery current consumption for 
		each attack, we determined what the (theoretical) maximum battery 
		operation time could be. Note, however, that in reality, the actual 
		operation time would be much lower than these due to the Peukert's 
		law~\cite{peukert}.}
	\label{tab:lifetime}
\end{table*}

{\bf Verifying wake-up functionality in various vehicles.}
Using the OBD-II device, we also verified how different vehicles 
(OEMs/years/models) react to the injection of a wake-up message. 
To confirm that the wake-up functionality exists in different cars, we chose 
various types of test vehicles: (compact and mid-size) sedans, coupe, 
crossover, PHEV, SUVs, truck, and an electric vehicle with model-years 
2008--2017.

Fig.~\ref{fig:wakeup_eval} shows how many distinct messages were observed on 
the CAN bus 1) when the ignition was on and 2) when we woke up ECUs on the bus 
via a wake-up message injection while the ignition was off.
Since the feasibility of wake-up stems from how the in-vehicle 
network standard is specified and thus implemented, i.e., instead of 
OEMs' design decisions, we have chosen not to identify/reveal the particular make 
and model used in our evaluation. 
Note, however, that the 11 examined test vehicles (shown in 
Fig.~\ref{fig:wakeup_eval}) are from different OEMs and also represent 
different models. For some vehicles, since their OBD-II pinout was configured 
to not provide full access to all of their buses, we only show those that were 
awakened in the accessible bus(es).

When waking up ECUs in some old cars (most with low-level trims), far less 
distinct message IDs and lower percentages of them (compared to the case with the 
ignition on) were observed on the bus than other newer cars. Since the number of ECUs is 
proportional to the number of distinct message IDs---although it is not linearly proportional---%
we can infer that there were less awakened ECUs transmitting them in older cars; 
not all messages can be sent by a single ECU due to the overhead.
This would most probably be due to the fact that the older cars 
do not require/provide any (or not many) standby functions (e.g., PKES).
For example, no standby functions (e.g., keyless entry, hands-free trunk 
opening) were 
provided in the 2008 and 2013 model-year test cars; the reason why  
none and only one ECU was awakened, when a wake-up message was injected on the bus.

On the other hand, when we injected a wake-up message on the buses of nine
2015--2017 model-year test vehicles, we observed that 49.12--94.95\% 
(75.44\% on average) of the distinct message IDs sent while the ignition was on, were 
{\em also} sent when ECUs were awakened while the ignition was off.
Such a high number/portion of ECUs being awakened from the wake-up message and 
thus sending more message IDs on the bus is because they had numerous 
standby functions installed for enhanced driver's experience and convenience 
(e.g., PKES/RKE, hands-free trunk opening, anti-theft)---a trend that is expected to expand.
These results corroborate the fact that vehicle ECUs are indeed equipped with 
the wake-up functionality (adhering to the standard) and can thus be exploited 
by the CAN adversary as an attack vector.

\begin{figure}[!t]
	\psfull \centering
	\includegraphics[width=0.6\linewidth]{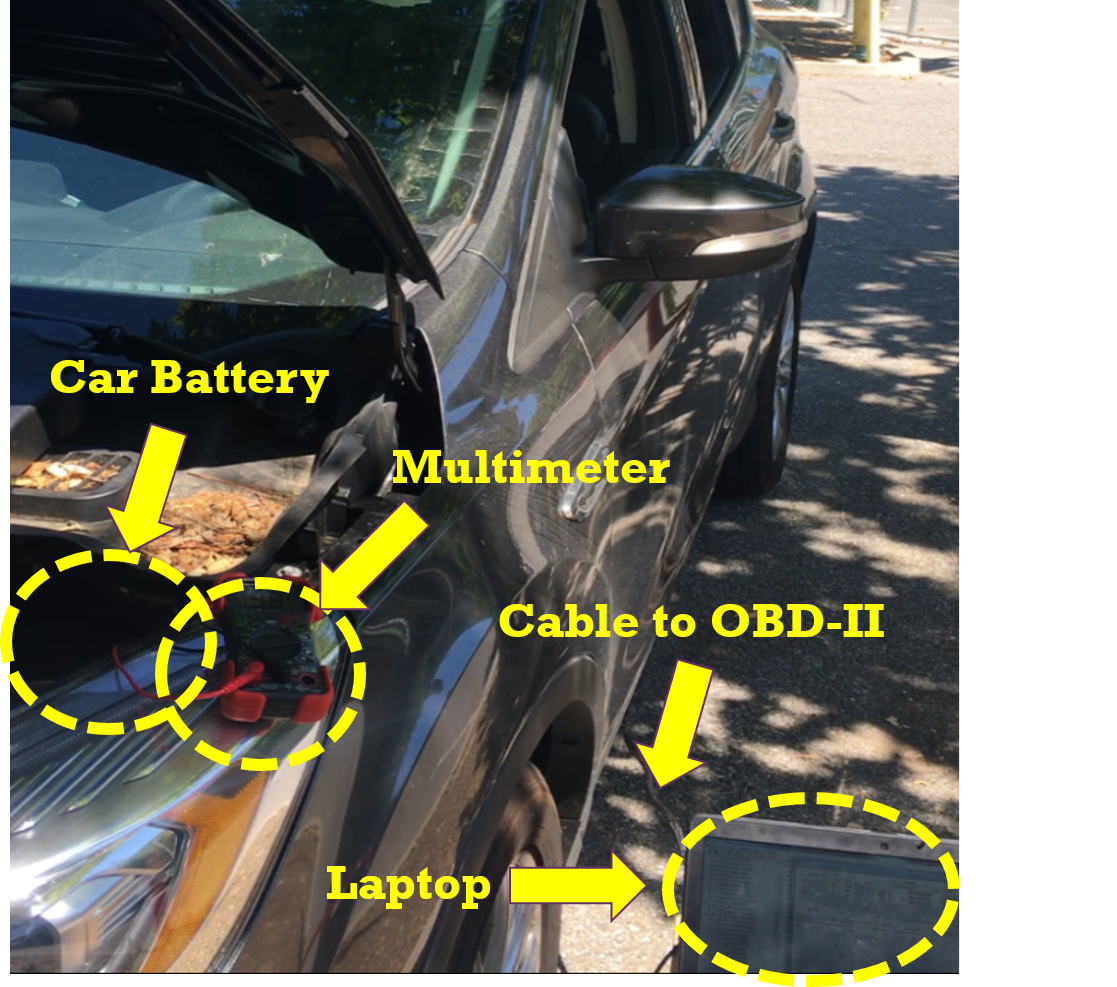}
	\caption{{\em Setup for measuring battery drain/consumption.}
		The discharge current from our test vehicle's 
		battery was measured using a multi-meter. }
	\label{fig:measure}
\end{figure}

%\begin{figure*}[!t]
%	\psfull \centering
%	\includegraphics[width=\linewidth]{consumption.png}
%	\caption{{\em Battery current consumption under different battery drain 
%			attacks.} We compared how many distinct message IDs were observed 
%			when the 
%		ignition was off but the ECUs were awakened via a wake-up message and 
%		when the ignition was on.}
%	\label{fig:consumption}
%\end{figure*}

\subsection{Battery-Drain Attack}
After verifying that the ECUs in our test vehicles can be awakened via 
basically any wake-up message (even with all 1s in ID, DLC, and DATA fields), 
we mounted 4 different types 
of {\atta}
on one of our test vehicles: 1) simple wake-up, 2) power mode control, 3) 
repetitive door unlock \& lock, and 4) opening the trunk.
We were able to reverse-engineer the control messages for those functionalities 
via the proposed driver-context-based reverse-engineering.

In order to measure the drained/discharged current from the car battery, we 
disconnected the negative cable from the negative battery terminal and connected 
our multimeter in series to the battery, i.e., one probe to the negative cable and 
the other to the battery terminal. We conducted the battery current draw test 
from the negative side to prevent accidental shorting and while the vehicle was 
parked with its ignition off. Then, as shown in Fig.~\ref{fig:measure}, we 
injected (iterative) sequence of control messages to the vehicle through the 
OBD-II port. 

{\bf Drained current.}
Table~\ref{tab:lifetime} summarizes how much current (on average) was measured 
to be drawn from the vehicle battery when each attack was mounted additionally. 
When the ignition was off and all ECUs were either asleep or completely off, 
i.e., when we did not wake up any ECU, the consumed current was 12.2mA. 
As expected, this value was below the parasitic drain threshold, which  
is about 30mA. However, when we just woke up the ECUs, the 
discharged current exceeded that threshold and drained 42mA. 
Exceeding the parasitic drain threshold even slightly is considered to
be a serious/critical problem.
As we controlled more functions that indirectly illuminated exterior/interior lights,
the average battery consumption increased further to 74.5, 101.1, and 153.3mA. 

{\bf More worse cases.}
It is important to note that, depending on the vehicle model, there could/would 
be (much) more of such controllable functions, thus allowing the attacker to 
drain the battery more quickly and easily. 
However, we only show 4 controls as examples for the proposed {\atta} 
since they are already very critical.
%in determining its criticality. 
Moreover, if the attacks are launched at 
night time when the exterior brightness is low, the car headlights will always 
come on (as one type of a welcome light) when the car is unlocked. 
As a result, the current drain will further increase sharply, i.e., 
drain the battery very quickly. However, in order to show the {\em minimum 
drain/discharge}, i.e., the {\em worst} possible case for the adversary, we 
conducted all our experiments outdoor during day time.

{\bf Expected battery operation time.}
In order to better understand how the increased battery current consumption 
relates to how quickly the attacker can immobilize the car, we determine the 
{\em battery operation time} as follows. 
Consider a 45Ah battery, which is the standard car battery capacity, with a 
State-of-Charge (SoC) of 70\%---the average battery SoC of a passenger 
vehicle---when parked. Then, since the minimum SoC for a cold start is 
considered to be 50\% in the {\em worst} case~\cite{wakeuptutorial}, i.e., 
worst-case for the adversary, our test vehicle's battery can theoretically
remain idle/parked for up to $\frac{45Ah \times 
\left(0.7-0.5\right)}{12.2mA}=737.7$ 
hours $\approx30.7$ days; something that is normally expected. Note that this is 
the theoretically maximum battery operation time since we considered the 
worst possible case for its derivation.

One can see from Table~\ref{tab:lifetime} how the maximum battery operation 
time was reduced as different types of {\atta}s were mounted. Theoretically, with 
the 4 control functions, it only takes a maximum of 2.44 days for an adversary 
to immobilize the vehicle via a battery-drain attack, i.e., can be achieved 
over the weekends. It is important to note, however, that it could take much 
shorter especially when the temperature is low and the battery is aged. 
More interestingly, in reality, the {\em actual} battery operation time is 
known to be much {\em shorter} than the theoretical/ideal value, i.e., much 
shorter than the times shown in Table~\ref{tab:lifetime}.
According to the Peukert's Law, because of intrinsic losses and 
the Coulombic efficiency being always less than 100\%, the actual battery 
operation time is much lower than the ideal/theoretic value in which the latter 
assumes the battery to be ideal~\cite{peukert}.
In fact, since the intrinsic losses in the battery escalate as load increases, 
the battery capacity is known to drop sharply as the 
drained/discharged current increases. This implies that as the adversary 
controls more functionalities, s/he can not only increase the battery 
consumption but also decrease the available battery capacity at the same time!
So, s/he can significantly reduce the battery operation time via a 
{\atta}, thus crippling the vehicle quickly, perhaps overnight.

\begin{figure}[!t]
	\psfull \centering
	\includegraphics[width=0.7\linewidth]{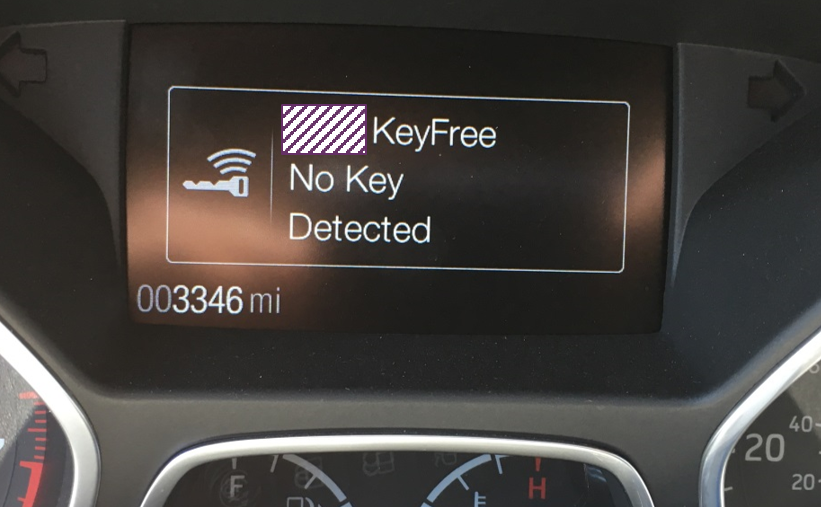}
	\caption{{\em Consequence of a DoB attack.} After the attack, the vehicle 
	could no longer detect that the key fob was inside it.}
	\label{fig:dobattack}
\end{figure}

\subsection{Denial-of-Body-control (DoB) Attack}
Through experiments on one of our test vehicles, we also verified the 
consequences of the proposed DoB attack.
Using the connected OBD-II device, we mounted a DoB attack as described in 
Section~\ref{sec:dobattack}.

After launching the DoB attack on one of our parked test vehicles, taking only 
a few seconds, we confirmed from the CAN traffic that all ECUs on the 
bus were continuously incurring and/or experiencing errors, causing all the 
ECUs to enter the bus-off state. After mounting the DoB attack, we observed 
most, but {\em not all} of the ECUs recovered from the bus-off state as 
configured. 
We observed that the number of distinct message IDs sent on the bus  
was actually reduced by 6 after the DoB attack.
By mapping those missing IDs to the actual transmitter ECU using the DBC file, 
we found that the RCM (Remote Control Module) did {\em not} recover from 
the bus-off, i.e., remained shut down, most probably due to its distinct recovery policy 
configuration (perhaps for anti-theft/engine-immobilizer purposes). 
Since the RCM was 
indefinitely off, the key fob was not authenticated and thus could not 
establish a connection to the vehicle. As shown in 
Fig.~\ref{fig:dobattack}, the vehicle could not detect that the key was in 
its vicinity; the key was in fact placed right in front of the dashboard. 
This consequence of the proposed DoB attack was reproducible on our test 
vehicle.

Of course, remote key fobs are now equipped with RFID chips that can be 
used for authentication, connection establishment, and thus starting the 
vehicle in case of a dead key fob battery. However, since the communication 
between the key fob's RFID and the vehicle was also configured to be governed 
by the RCM, the RFID-based (emergency) start did not work either.
More interestingly, after mounting the DoB attack, when we tried to open the 
doors or trunk to enter the car, we could not since the RCM was not 
functioning and thus failed to authenticate the key fob. 
The only way to get in was actually using the back-up physical key hidden in 
the key fob. Note, however, that the car did not start anyway (as shown in 
Fig.~\ref{fig:dobattack}) even though we were able to enter it!

As discussed earlier, this consequence comes from the fact that OEMs (or 
their ECUs) may have different bus-off recovery configurations. In our test 
vehicle, the setting of an RCM to not recover from the bus-off ``favors'' the 
attacker 
in mounting a critical DoB attack. We found the only way to restore the 
vehicle back to its original state after a DoB attack was to disconnect the 
battery, wait for a few minutes, and re-connect the battery. 
Such a process resets the {\em states} stored in each ECU and thus 
lets them run in their original/intended states. However, imagining a victim confronting 
the symptoms of DoB attack, i.e., the key fob neither working nor being 
detected, s/he might first try to change the key fob battery. Obviously, since 
that won't work, s/he would consider the car battery completely dead 
and therefore, would probably have the car towed to the service station for a battery
replacement, wasting money and time unnecessarily!
\section{Related Work}\label{sec:relworks}
%Exploiting vulnerabilities in in-vehicle networks, researchers have proposed 
%various ways on how to mount attack vehicles and thus control/compromise them.
Exploiting a remotely compromised ECUs, researchers have shown how various 
vehicle maneuvers (e.g., braking, steering) can be (maliciously) controlled by 
injecting packets into the in-vehicle network~\cite{koscher,miller}.
Similarly, in 2015, researchers were able to compromise and remotely 
kill a Jeep Cherokee running on a highway~\cite{miller3}, which triggered a 
recall of 1.4 million vehicles. In 2016 and 2017, researchers were able to hack 
a Tesla model S and model X, respectively, exploiting software vulnerabilities, 
and controlled vehicle maneuvers~\cite{teslahack}.
Researchers have also demonstrated that an adversary can also shut down a 
specific ECU or even the entire in-vehicle network merely via packet injection~\cite{busoff}.
The authors of \cite{tpms} also succeeded in a remote 
attack via a vehicle's tire pressure monitoring system (TPMS).
Researchers also proposed new hardware which can generate/fabricate 
magnetic fields, spoof the wheel speed sensor on a vehicle, and thus activate 
the Anti-lock Braking System (ABS)~\cite{ABSattack}.

Although such attacks were effective, they were mounted and thus 
considered malicious only when the vehicle is running. In contrast, 
we focused on, and proposed new attacks that are effective 
when the ignition is {\em off}. Moreover, the controlled functionalities had 
to be very different from existing attacks since those that are considered ``critical'' are 
different. The steering being maliciously controlled is clearly
safety-critical when the vehicle is running, but not so when the vehicle is 
parked; illumination of exterior/interior lights might be a more critical problem!

\section{Discussion}\label{sec:disc}
{\bf Countermeasures.}
As the proposed {\atta} and {\attb} are mounted with the ignition off,
the design principles of their countermeasures have to be very different from 
the state-of-the-art defenses, which are mostly concerned with attacks on ``running'' vehicles.
For example, as of the current CAN standard, since the wake-up itself can be 
achieved with any CAN message having a 010 bit-sequence, adding MAC 
or message encryption cannot prevent the adversary from waking up ECUs; a 
message with MAC/encryption will still have such a sequence.

The cornerstone of {\atta} and {\attb} is to wake up ECUs asleep on the bus 
while the ignition is off. So, as their feasible and effective 
countermeasure, one can think of continuously running an Intrusion Detection 
System (IDS) even when the ignition is off in order to capture any abnormal 
wake-up messages;  wake-up messages usually  should not be seen very frequently. 
However, since the operation of an IDS would increase the 
current drawn from the battery, such an approach may defeat the very purpose
of reducing battery consumption. % and must thus be carefully examined. 
Like other ECUs asleep, the IDS ECU can also be configured to sleep most of the time 
%(to minimize battery drain) 
and wake up only when it sees a wake-up message. 
In such a case, as a countermeasure against both types of attack, 
the wake-up pattern of an IDS can be modeled and used to detect any abnormal 
wake-up requests on the bus without continuously running it. 
Similarly, the IDS can be configured to wake up periodically, check the 
battery SoC---if there was any significant drain recently---and  
react accordingly. Moreover, especially for the DoB attack, how to recover from 
the bus-off state has to be re-examined in order to prevent the consequences 
of the DoB attack, as we had demonstrated.

{\bf Enhanced wake-up functionality.}
Partial networking---i.e., partial deactivation of subnets within a given 
network---has been discussed and planned to be installed by car 
manufacturers. This is to reduce energy consumption and thus CO$_2$
emissions~\cite{wakeuptutorial2}. In such a setting, only the
pre-defined wake-up messages that pass the wake-up masks/filters of selective 
ECUs can wake up those ECUs during operation. However, since that message is 
``pre-defined'' and can easily be learned by observing the CAN traffic and its 
sudden change in the number of message IDs (as in Section~\ref{sec:reverse}), 
the wake-up message itself can still be learned and used by an 
adversary. In fact, the introduction of partial networking will increase the 
number of ECUs to sleep rather than being completely turned off, and thus 
provide a larger attack space for mounting the proposed attacks.

{\bf Limitations.}
We considered/assumed that the (CAN) adversary has access to the in-vehicle 
network via a compromised ECU (either a compromised OBD-II dongle or an 
in-vehicle ECU)---a limitation of our approach---to mount the proposed attacks. 
However, we must not overlook the fact that the adversary might not 
even need a compromised ECU to mount the proposed attacks. 
To further enhance the driver's convenience, companies such as 
Volvo~\cite{volvo_keyless}, Lexus~\cite{lexus_keyless}, and Tesla~\cite{tesla_keyless} 
started to let car owners unlock/lock their cars by using their smartphone apps with an 
eventual goal to totally replace key fobs with smartphone software. 
This means the existence of another (new) attack vector for mounting  {\atta}: compromising those apps! 
As more of such vehicle-related technologies evolve and get deployed, there 
may be more (intelligent) ways of mounting the proposed attacks.

Although we succeeded in mounting {\atta} and {\attb} on our 2017 model-year test vehicle, 
not many ECUs were awakened when a wake-up message was injected
in {\em older} vehicles, because they had less
standby functions than {\em newer} models, and 
thus had less ECUs asleep while the ignition was off.
The proposed attacks will likely be easier and more effective to be mounted on newer models 
as we observed in waking up more ECUs in newer vehicles (Fig.~\ref{fig:wakeup_eval}).
For the \attb, however, since its success/feasibility will totally depend on 
how the OEMs configured their ``bus-off recovery'' for different ECUs, it might 
not be as feasible as {\atta}. The battery-drain attack will still be feasible 
unless the standard wake-up procedure is changed or standby functions are not 
installed. % \kt{(unpleasantly) hindering driver's convenience.}
\section{Conclusion}\label{sec:conclusion}
In this paper, we discovered two new important vulnerabilities in vehicle 
availability:  {\em battery-drain} and {\em Denial-of-Body-control} (DoB) attacks.
They are counter-intuitive in that attacks are commonly believed to be possible and effective only while the ignition is on. 
Specifically, we have shown that an attacker can wake up ECUs on the bus, even while the ignition is off, 
mount the proposed attacks, and then immobilize a parked vehicle with its ignition off. 
Through extensive experiments on real vehicles, we showed 
that such attacks are indeed easy to mount and very critical to vehicle availability.
Ironically, the adversary exploits, as attack vectors, the in-vehicle network features originally 
designed for either energy-efficiency (e.g., simple wake-up signals) or enhanced user/driver 
experience/convenience (e.g., standby functions).
There could still remain different types of unknown and unintuitive vehicle vulnerabilities, like the two proposed here. 
It is therefore important to analyze and understand what consequences 
existing built-in/standardized functionalities can lead to. This calls for concerted
efforts from both academia and industry on this possibility and countermeasures thereof in order to build secure vehicles.

\bibliographystyle{IEEEtran}
\bibliography{referenceAll}
%\theendnotes

%\input{Supplement}
\end{document}